# CRITICAL CURRENT DISTRIBUTION IN SPIN TRANSFER SWITCHED MAGNETIC TUNNEL JUNCTIONS

Mahendra Pakala, Yiming Huai, Thierry Valet, Yunfei Ding and Zhitao Diao

*Grandis Inc., 1266 Cadillac Court, Milpitas CA 95035*

The spin transfer switching current distribution within a cell was studied in magnetic tunnel junction based structures having alumina barriers with resistance-area product (*RA*) of 10 to 30 $\Omega$–$\mu$m$^2$ and tunneling magneto-resistance (*TMR*) of ~20%. These were patterned into current perpendicular to plane configured nano-pillars having elliptical cross-sections of area ~0.02 $\mu$m$^2$. The width of the critical current distribution (sigma/average of distribution), measured using 30 ms current pulse width, was found to be 7.5% and 3.5% for cells with thermal factor ($K_uV/k_BT$) of 40 and 65 respectively. The distribution width did not change significantly for pulse widths between 1 s and 4 ms. An analytical expression for probability density function, $p(I/I_{co})$ was derived considering the thermally activated spin transfer model, which supports the experimental observation that the thermal factor is the most significant parameter in determining the within cell critical current distribution width.

Since the observation of spin transfer (ST) induced magnetization switching in metallic systems [1], ST switching has stimulated considerable interest in recent years due to its rich fundamental physics and potential for application in devices such as non-volatile memory (advanced MRAM) and high frequency oscillators [2]. For non-volatile memory application, spin transfer switching of magnetic tunneling junction (MTJ) is of particular interest because of the high read signal and high cell resistance possible compared to pure metallic systems. However there is limited work reported in this area. Clear observation of ST switching in MTJ was reported by Y. Huai et. al. (at room temperature) [3] and later by Fuchs et. al. (at 77 K) [4]. Low switching current density values (~$10^5$ A/cm$^2$), which is required for compatibility of the MTJ with underlying transistor, and a tight distribution of switching current density (within a cell and within a die sigma of less than 3%), which is required to get adequate margins between read and write currents, are two very important parameters for realizing ST writing scheme in functional devices. In this paper, we discuss the observed switching current distribution within a cell for MTJ's with *RA* ranged from 10 to 30 Ω–μm$^2$ and with *TMR* values of ~20%. The implication of thermal stability of bits is discussed both in terms of the stability of memory cells in idle state as well as the behavior of memory cell during read/write operations.

MTJ structures (Ta5/PtMn20/CoFe2/Ru0.8/CoFeB3/$Al_2O_3$/CoFeB1/NiFe2/CoFeB1/Cu5/CoFe2/PtMn15/Ta5 in nm) were deposited using a Singulus PVD cluster system and annealed at 250-300$^o$C for 2 hours in a magnetic field of 1 Tesla. The thin tunneling barrier was formed by natural oxidation of an Al layer in a pure oxygen atmosphere. The MTJ films were subsequently patterned into deep sub-micron ellipse shaped pillars using e-beam patterning with etch-back planarization of insulating layer [5]. A quasi-static tester with pulse current capability was used to measure resistance ($R$) as function of magnetic field ($H$) and current ($I$) at room temperature. The switching current distribution was measured by repeating the $R$ vs. $I$ measurements for >20 times in an external applied magnetic field, $H_a$, equal to the offset field, $H_{off}$, experienced by free layer. Representative resistance ($R$) versus applied field ($H$) and resistance versus applied current ($I$) are shown in Fig. 1 for two cells on the same wafer. While the first data set: (a) and (b) are for a site (cell) with coercivity ($H_c$) of 25 Oe, the second set of data, (c) and (d) are for a site with $H_c$ of 48 Oe. The distribution of switching current, for samples shown in Fig. 1(b) and (d) is more clearly seen in Fig. 2, where histograms are plotted for parallel to anti-parallel ($I_{c\ p\text{-}ap}$) and anti-parallel to parallel ($I_{c\ ap\text{-}p}$) switching currents for the two sites. The switching currents of Fig. 2 were obtained from $R$ vs. $I$ plots using 30 ms pulse width. Figs. 2(a) and (b) correspond to the first site and Figs. 2(c) and (d) correspond to the second site. As seen in Fig. 4, the actual distribution is non-symmetric, however the distribution can be approximated by Gaussian form in order to compare the distribution width. The switching current distribution (sigma of $I_{c\ a\text{-}ap}$ and $I_{c\ p\text{-}ap}$, divided by the average) is significantly larger for the first site (7.5%) compared to the second site (3.5%). This is also true for histograms at current pulse widths of 1 s, 300 ms and 4 ms, representing a total of 100 switches. From the cell dimensions (180 nm x 140 nm for first site and 190 nm x 130 nm for second site) and assuming 20 Oe induced anisotropy, the total in-plane uni-axial anisotropy of the cells can be calculated for two aspect ratios used. Then from these values, the thermal factor ($K_uV/k_BT$) was estimated for the two cells to be 40 and 65 respectively at T= 300 K. Despite the use of pulsed current, some degree of heating of the MTJ is expected above the room temperature, however the results do not change significantly.

The critical switching current distribution data for a site measured at different pulse widths is shown in Fig. 3. Fig. 3(b) was used to estimate the value of $I_{co}$ (~ the extrapolated value of $I_c$ at 1 ns pulse width). The distribution of switching current (envelope of histogram) was plotted as a function of $I_c/I_{co}$ in Fig. 3(a). From the data, we observe that the width of the distribution does not change significantly for the different pulse widths, however the average $I_c/I_{co}$ value shifts toward higher values for smaller pulse widths. This indicates that at smaller pulse widths, the effect of thermal activation is reduced. In order to understand the effect of the thermal factor ($K_uV/k_BT$) of cells on the width of critical current distribution, as well as the effect of pulse width on the average critical current, we derive an analytical expression based on thermally activated model for spin transfer switching [6-10]. The expression can be used to directly relate device level measurements (within cell current distribution) to the physical parameters mentioned above.

Consider an ensemble of magnetic moments, with $m_x$ being the fraction of the ensemble aligned in the positive x-axis. For complete magnetization alignment (parallel) in the positive $x$ direction, $m_x$ takes the value of +1. Then the rate equation can be written starting from a parallel (p) state as

$$\frac{d}{dt}\langle m_x\rangle = \left[1-\langle m_x\rangle\right]\tau^{-1}_{\mathrm{ap}\to\mathrm{p}} - \langle m_x\rangle \tau^{-1}_{\mathrm{p}\to\mathrm{ap}} \quad (1)$$

where, $\tau_{p\to ap}$ is the relaxation time with the ensemble in an initial parallel state, given by

$$\tau_{p\to ap} = \tau_0 \exp\left[\frac{K_u V}{k_B T}\left(1+\frac{H(t)}{H_K}\right)^2\left(1-\frac{I(t)}{I_{c0}}\right)\right] \quad (2)$$

$H(t)$ is the applied field and $I(t)$ is applied current as a function of time. $I_{co}$ is the zero temperature switching current, as given in Reference 6. A similar expression for $\tau_{ap\to p}$ can be shown. Under an applied current that favors the parallel to anti-parallel transition, the rate equation can be simplified to include parallel to anti-parallel term only, since the $\tau_{ap\to p}$ would be very large. Finally for the measurement method where a pulse of constant current is applied for duration of $t_p$, one can integrate the rate equation over time zero to $t_p$ and obtain an expression for $m_x$ as a function of $t_p$.

$$\langle m_x\rangle (t_p) = \exp\left[-\frac{t_p}{\tau_{p\to ap}}\right] \tag{3}$$

This is the cumulative probability distribution as a function of $t_p$. A formal differentiation of this expression with respect to current would give the probability distribution function of switching current with respect to $I$.

$$p\left(\frac{I}{I_{c0}}\right) = \frac{K_u V}{k_B T}\frac{\left(1+\frac{H(t)}{H_K}\right)^2}{I_{c0}}\frac{t_p}{\tau_{p\to ap}}\exp\left[-\frac{t_p}{\tau_{p\to ap}}\right] \tag{4}$$

Plots of this expression are shown in Fig. 4. In Fig. 4 (a), the effect of the thermal factor ($K_uV/k_BT$) on the distribution width is shown, while in Fig. 4(b) the effect of pulse width is shown. The distribution width obtained from the analytical expression (half width at 1/e of the height of peak divided by the average) was found to be 7% and 3% for the thermal factors of 40 and 65 respectively.

From Fig. 4(a), it is seen that the width of the distribution decreases significantly with increasing $K_uV/k_BT$. Further, as seen from Fig. 4(b), the average $I/I_{co}$ value of ~0.52 is derived for a pulse width of 0.1 sec, which is close to the experimental value for similar pulse width (0.3 s) shown in Fig. 3(a). Thus good agreement between the analytical expression and measured data is observed. Also from Fig. 4(b), we see that though the width of the distribution does not change significantly with pulse width, however, the average value of critical current increases with decreasing pulse width because the effect of thermal activation becomes more significant for longer pulse widths.

In summary, it was found experimentally as well as analytically that the thermal factor ($K_uV/k_BT$) is the most dominant factor in determining the within cell critical current distribution width. In spin transfer writing scheme for advanced MRAM, the margin in both read and write operations is influenced by the thermal factor, via its effect on switching current distribution. To get sufficient separation between read current, write current and the barrier breakdown current, a tight distribution of the write current is required and hence the need for a high thermal factor. Also a high thermal factor reduces the probability of unintentional reversal of a stored bit through thermally activated process. Therefore, a high thermal factor or $K_uV/k_BT$ is required, for both read and write reliability, as well as for long-term retention of the written data.

We would like to thank Y. Higo and others from Sony Corp. for the helpful discussions, Dr. Wolfram Mass, Dr. Berthold Ocker and Dr. Juergen Langer of Singulus Inc. for joint development of low RA MTJ films and, Dr. Alex Panchula and Dr. Dmytro Apalkov of Grandis for useful suggestions.

[1] F. J. Albert, J. A. Katine, R. A. Buhrman and D. C. Ralph, Appl. Phys. Lett., **77**, 3809, (2000), J. A. Katine, F. J. Albert, R. A. Buhrman, E. B. Myers and D. C. Ralph, Phys. Rev. Lett., **84**, 3149 (2000).

[2] J. Z. Sun, D. J. Monsma, D. W. Abraham, M. J. Rooks, and R. H. Koch, Appl. Phys. Lett., **81**, pp. 2202-2204 (2002); M. R. Puffall, W. H. Rippard, and T. J. Silva, *ibid.*, **83**, pp. 323-325 (2003); S.

Urazhdin, H. Kurt, W. P. Pratt, and J. Bass, *ibid.*, **83**, pp. 114-116; F. B. Mancoff, R. W. Dave, N. D. Rizzo, T. C. Eschrich, B. N. Engel and S. Tehrani, *ibid.*, **83**, 1596-1598 (2003); Y. Jiang, T. Nozaki, S. Abe, T. Ochiai, A. Hirohata, N. Tezuka, and K. Inomata, Nature Materials, **3**, 1, (2004).

[3]Y. Huai, F. J. Albert, P. Nguyen, M. Pakala and T. Valet, Appl. Phys. Lett. **84**, 3118, (2004).

[4]G. D. Fuchs, N.C. Emley, I. N. Krivorotov, P. M. Braganca, E. M. Ryan, S. I. Kiselev, J. C. Sankey, D. C. Ralph, R. A. Buhrman and J. A. Katine, Appl. Phys. Lett., **85**, 1205, (2004).

[5]Y. Ding, M. Pakala, P.Nguyen, H. Meng, J. P. Wang and Y. Huai, to be published in J. Appl. Phys., 2005

[6]R. H. Koch, J. A. Katine, and J. Z. Sun, Phys. Rev. Lett., **92**, 088302 (2004)

[7]D. M. Apalkov and P. Visccher, arXiv:cond-mat/ 0405305 v2 1 Jun 2004

[8] Z. Li and S. Zhang, Phys. Rev. **B69**, 134416 (2004)

[9]E. B. Myers, F. J. Albert, J. C. Sankey, E. Bonet, E. Bonet, R. A. Buhrman and D. C. Ralph, Phys. Rev. Lett., **89**, 196801 (2002).

[10]I. N. Krivorotov, N. C. Emley, A. G. F. Garcia, J. C. Sankey, S. I. Kiselev, D. C. Ralph and R. A. Buhrman, Phys. Rev. Lett, **93**, 166603 (2004)

LIST OF FIGURES

Fig. 1. Distribution of critical current within a cell for cell designed with $K_uV/k_BT$ of 40 in (b) and 65 in (d). The *R* vs. *I* loops were measured at an applied field equal to the offset field, $H_{off}$. The respective *R* vs. *H* loops are shown in (a) and (c).

Fig. 2. Histogram of critical current distribution, at 30 ms pulse widths, for parallel (p) to anti-parallel (ap) as well as ap to p state magnetization switching. (a) and (b) correspond to cell with $K_uV/k_BT$ of 40 while (c) and (d) correspond to cell with $K_uV/k_BT$ of 65.

Fig. 3 Distribution of critical current within a cell (a) and the average critical current (b) as a function of pulse widths (in sec).

Fig. 4. Plot of equation (4). In (a) effect of thermal factor $K_uV/k_BT$ is plotted keeping $t_p$ constant at 0.1s, while in (b) effect of $t_p$ is plotted keeping $K_uV/k_BT$ constant at 40.

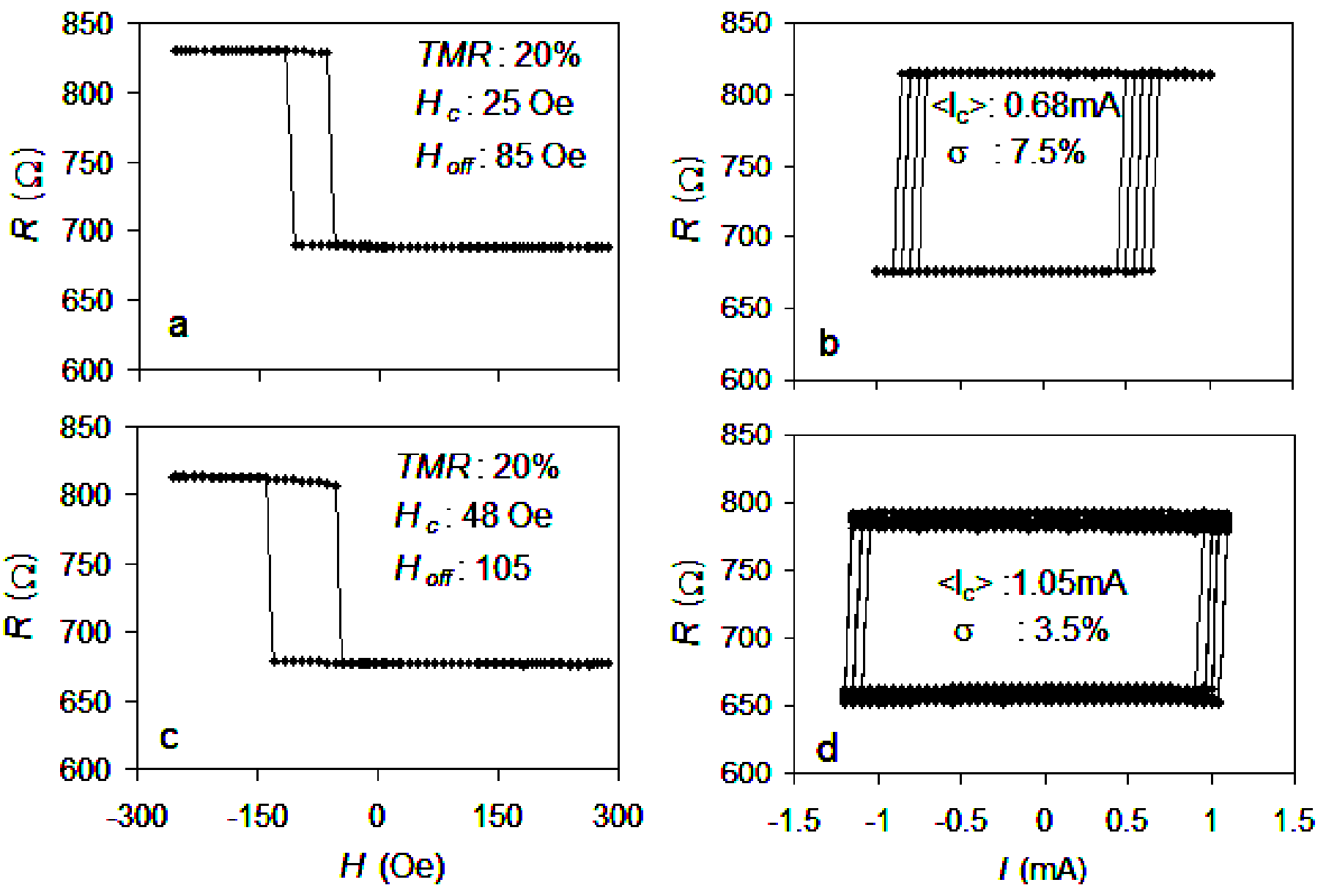

Fig. 1, M. Pakala et. al.

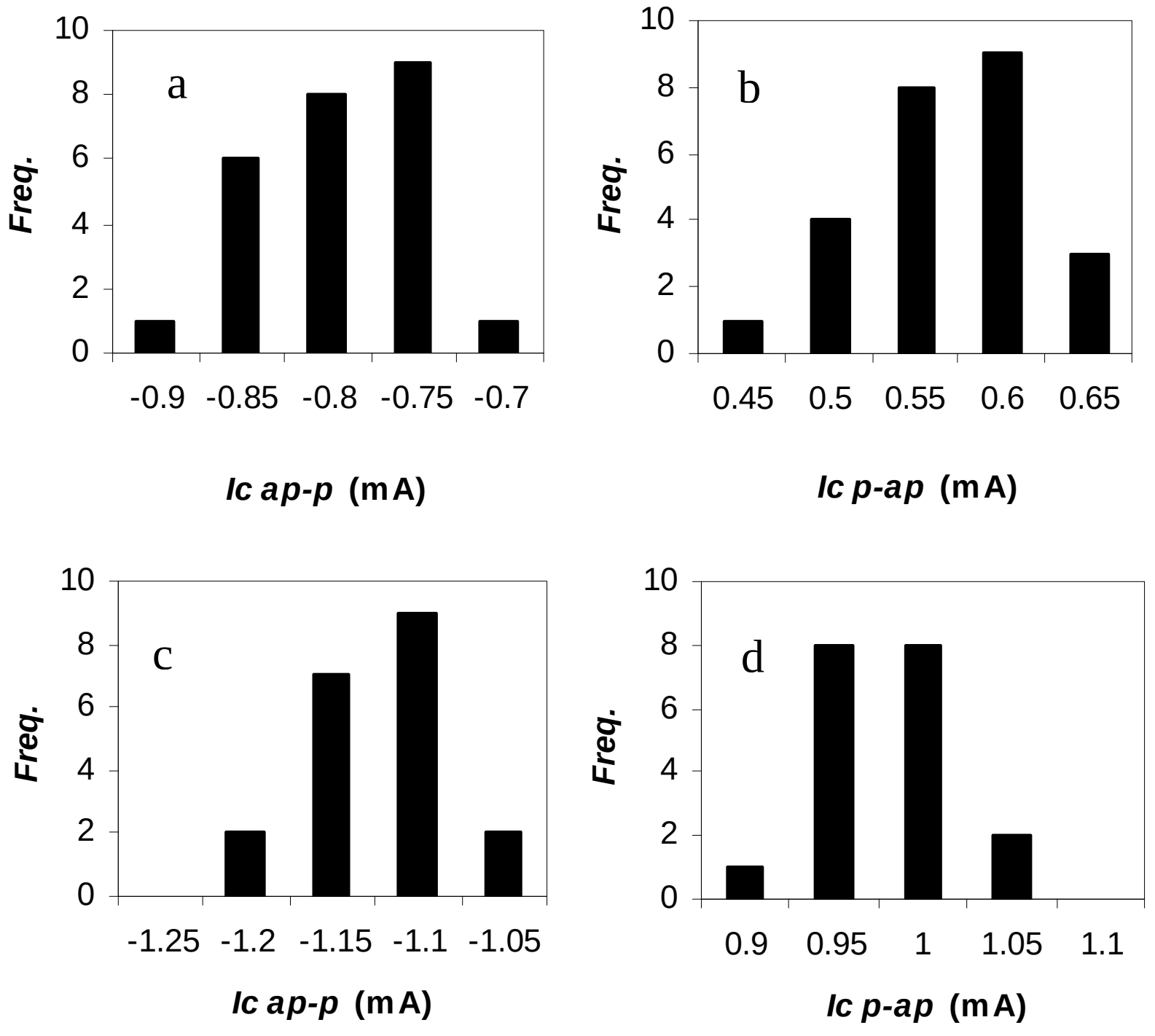

Fig. 2. M. Pakala et. al.

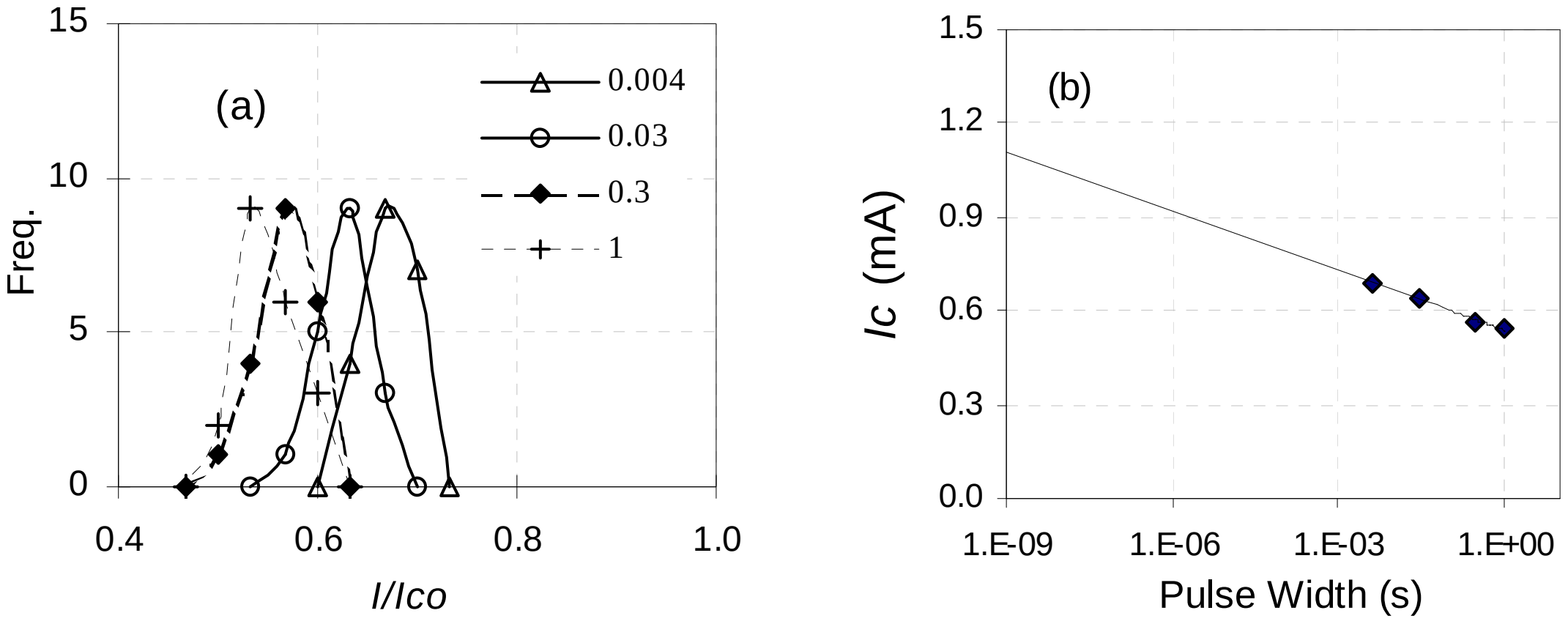

Fig. 3, M. Pakala et. al.

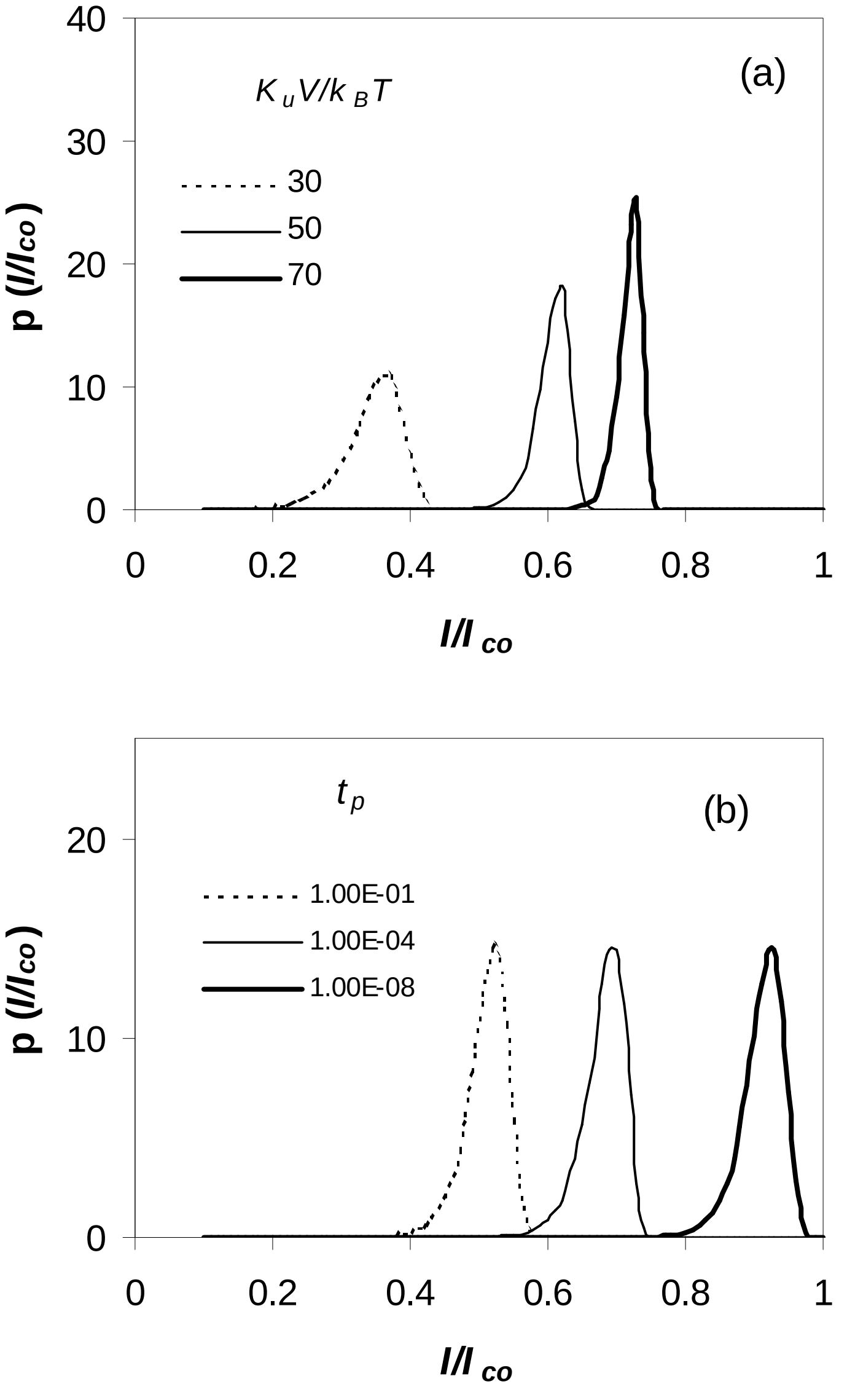

Fig. 4, M. Pakala et. al.